\documentclass[prd,showpacs,showkeys,preprintnumbers,amsmath,amssymb,floatfix]{revtex4}

\usepackage[utf8]{inputenc}
\usepackage{hyperref}

\usepackage{dcolumn}
\usepackage{bm}
\usepackage{graphicx}
\usepackage{subfigure}

\def\be{\begin{equation}}
\def\ee{\end{equation}}
\def\bestar{\begin{equation*}}
\def\eestar{\end{equation*}}

\begin{document}

\title{Black holes and gravitational waves sourced by non-linear duality rotation-invariant conformal electromagnetic matter} 

\author{Daniel Flores-Alfonso}
\email{daniel.flores@xanum.uam.mx}
\affiliation{Departamento de F\'{i}sica \\ Universidad Aut\'{o}noma Metropolitana - Iztapalapa\\
                   Av. San Rafael Atlixco 186, C.P. 09340, México City, M\'exico}         

\author{Blanca Ang\'elica Gonz\'alez-Morales}
\email{gombstar@xanum.uam.mx}
\affiliation{Departamento de F\'{i}sica \\ Universidad Aut\'{o}noma Metropolitana - Iztapalapa\\
                   Av. San Rafael Atlixco 186, C.P. 09340, México City, M\'exico}         
                                     
\author{Román Linares}
\email{lirr@xanum.uam.mx}
\affiliation{Departamento de F\'{i}sica \\ Universidad Aut\'{o}noma Metropolitana - Iztapalapa\\
                   Av. San Rafael Atlixco 186, C.P. 09340, México City, M\'exico}         
                   
\author{Marco Maceda}
\email{mmac@xanum.uam.mx}
\affiliation{Departamento de F\'{i}sica \\ Universidad Aut\'{o}noma Metropolitana - Iztapalapa\\
                   Av. San Rafael Atlixco 186, C.P. 09340, México City, M\'exico}

\date{\today}

\begin{abstract}
Maxwell's equations are invariant under both duality rotations and conformal transformations. Recently Bandos, Lechner, Sorokin, and Townsend have found
a nonlinear generalisation of electrodynamics
which possesses both of these symmetries.
We couple this theory to General Relativity and explore self-gravitating configurations
with a clear physical interpretation.
We find charged black holes and exact gravitational waves. The black hole geometry is Reissner-Nordstr\"om,
however, the non-linearity parameter acts as a screening factor allowing extremal configurations where the mass of the black hole is smaller than its charge.
Furthermore, we also discuss the phenomenon of birefringence and determine the optical metrics associated to the propagation of photons.
\end{abstract}



\maketitle

\section{Introduction}
\label{intro}

Non-linear electrodynamics (NED) has, by now, a long history in Physics. After the original non-linear Born-Infeld (BI) electrodynamics~\cite{Born:1934gh} leading to finite energy density for the electromagnetic field, it was quickly recognised that fundamental processes require a modification of Maxwell's electrodynamics in a non-trivial way. A clear example is Euler-Heisenberg non-linear electrodynamics~\cite{Heisenberg:1935qt}, an effective theory resulting from the polarisation of the vacuum where virtual charges surround the actual charges and currents~\cite{Ruffini:2013hia}. On the other hand, Lorentz invariance is a key symmetry of modern theoretical physics. The fact that Maxwell's equations
are invariant under Lorentz transformations
was the foundation of the principle of relativity.
Moreover, the limiting effect Lorentz transformations have on the velocity of massive particles is what motivated the Born-Infeld Lagrangian in the first place~\cite{Born:1934gh}. The nonlinear nature in the BI constitutive relation is what has lead to the study of NEDs in general. These are only a few examples of how geometry and field theory
enrich each other; for a further example
relating to conformal and duality-rotation symmetry
see~\cite{Boyer:1984wb}.

In General Relativity (GR), field singularities such as the ones the Born-Infeld theory removes are met with
more forgiving eyes. For example Reissner-Nordstr\"om
black holes have a field singularity in the same place where spacetime does. However, the singularity
is shielded by an event horizon. For the BI analogue
the field is regular everywhere but the black hole preserves its protected singularity. Nonetheless, it is NED which makes it possible to construct regular black holes~\cite{AyonBeato:1998ub}. This solution arises from Einstein theory coupled to nonlinear electrodynamics of Born-Infeld-type allowing for freedom of duality rotations~\cite{Salazar:1987ap}.

In this work, we consider gravitating configurations
in electrovacuum with a nonlinear costitutive relation and matter equations invariant under Hodge duality rotations and conformal transformations. This conformally extended duality-invariant theory was recently proposed in~\cite{Bandos:2020jsw}. Notwithstanding, a direct and simple derivation of the theory's Lagrangian is available in~\cite{Kosyakov:2020wxv2}. However, the Hamiltonian formalism (used in the original work) is at times more convenient, e.g., it clarifies that this nonlinear extension of standard electrodynamics is unique. Moreover, the Euler-Lagrange equations of the theory are non-analytical
when considering plane-waves. Whilst the Hamiltonian
equations remain well-defined. For the sake of brevity, henceforth we refer to this electromagnetic theory as ModMax. The focus of our analysis is on basic black hole and gravitational wave spacetimes. While doing so, we study the effect of allowing for the maximum number of symmetries which a NED may possess. Besides these features, when considering extensions of Maxwell's electrodynamics, it is always a natural question to ask if the phenomenon of birefringence exists~\cite{Balakin:1997dv}. As we know, a unique light-cone structure is intimately related to the absence of birefringence and in this regard, Born-Infeld electrodynamics has the peculiar feature of showing no birefringence. Since ModMax electrodynamics is a low-energy limit of a one-parameter generalisation of BI, as discussed in~\cite{Bandos:2020jsw}, we investigate if this phenomenon is present with the purpose of gaining further insight into the physical properties of gravitating ModMax systems.

This Letter is organised as follows: in Sec.~\ref{secc:2} we briefly review ModMax electrodynamics and its formulation in terms of Pleba\'nski's dual variables~\cite{Plebanski:1970}. We, then, construct spherically symmetric charged black holes which are solutions to GR coupled to ModMax in Sec.~\ref{secc:3}. We analyse electric, magnetic and dyonic configurations separately. In Sec.~\ref{secc:4}
we present type II and type N gravitational waves
sourced by ModMax matter.
Afterwards, we discuss the phenomenom of birefringence in Sec.~\ref{secc:5}. Lastly, we end with some final remarks and perspectives in the Conclusions section.

\section{ModMax electrodynamics and Pleba\'nski variables}
\label{secc:2}

We are interested in the ModMax electrodynamics which
is described by the Lagrangian~\cite{Bandos:2020jsw,Kosyakov:2020wxv2}
\be
L_{ModMax}(x, y) = -x\cosh \gamma + (x^2 + y^2)^{1/2}  \sinh \gamma
\ee
where
\be
x := \frac 14 F_{\mu\nu} F^{\mu\nu}, \qquad y := \frac 14 F_{\mu\nu}\tilde F^{\mu\nu},
\ee
are the electromagnetic Lorentz invariants in terms of the electromagnetic tensor $F_{\mu\nu}$ and its dual $\tilde F_{\mu\nu} := \frac 12 \epsilon_{\mu\nu\sigma\rho} F^{\sigma\rho}$. From the ModMax Lagrangian the Pleba\'nski dual variable reads~\cite{Plebanski:1970,Gibbons:2001sx}
\begin{eqnarray}
P_{\mu\nu} &:=& -L_x F_{\mu\nu} - L_y \tilde F_{\mu\nu}
\nonumber \\[4pt]
&=& \left[ \cosh \gamma - \frac x{(x^2 + y^2)^{1/2}} \sinh \gamma \right] F_{\mu\nu}
\nonumber \\[4pt]
&&-\frac {y \sinh \gamma}{(x^2 + y^2)^{1/2}} \tilde F_{\mu\nu}. \label{Pmodmax}
\end{eqnarray}
Its dual is
\begin{eqnarray}
\tilde P_{\mu\nu} &=& \left[ \cosh \gamma - \frac x{(x^2 + y^2)^{1/2}} \sinh \gamma \right] \tilde F_{\mu\nu}
\nonumber \\[4pt]
&&+\frac {y \sinh \gamma}{(x^2 + y^2)^{1/2}} F_{\mu\nu}. 
\end{eqnarray}
Using these expressions, we calculate the electromagnetic invariants
\begin{eqnarray}
s &:=& - \frac 14 P_{\mu\nu} P^{\mu\nu}
\nonumber \\[4pt]
&=& - x \cosh 2\gamma + (x^2 + y^2)^{1/2} \sinh 2\gamma, 
\nonumber \\[4pt]
t &:=& - \frac 14 P_{\mu\nu} \tilde P^{\mu\nu} = -y.  
\end{eqnarray}
Notice that the last relation is also true for the Born-Infeld electrodynamics Lagrangian; this is a nice feature of the ModMax electrodynamics as a weak-field limit of the generalised Born-Infeld structure introduced in~\cite{Bandos:2020jsw}. The previous relations may be inverted to obtain
\begin{eqnarray}
x &=& - s \cosh 2\gamma + (s^2 + t^2)^{1/2} \sinh 2\gamma,
\nonumber \\[4pt]
y &=& -t.
\end{eqnarray}
Due to its construction, the ModMax Lagrangian is dual to itself in the sense that 
\be
\hat L_{ModMax} := -\frac 12 P^{\mu\nu} F_{\mu\nu} - L_{ModMax}
\ee
satisfies $\hat L_{ModMax} = L_{ModMax}$ when written in terms of the $x, y$ variables; in terms of the $s, t$ variables, we have
\begin{align}
\hat L_{ModMax}(s, t) &= -x \cosh \gamma + (x^2 + y^2)^{1/2} \sinh \gamma
\nonumber \\[4pt]
&= -x\cosh \gamma + \frac {s + x \cosh 2\gamma}{\sinh 2\gamma} (s^2 + t^2)^{1/2}
\nonumber \\[4pt]
&= s \cosh \gamma - (s^2 + t^2)^{1/2} \sinh \gamma. 
\end{align}
This result is similar to the well-known relation between $L$ and $\hat L$ in non-linear Born-Infeld or Euler-Heisenberg electrodynamics. From these and other 
NEDs we have learned that the nature of the electromagnetic matter impacts the supporting geometry. For example, in black holes, NEDs induce 
charge screening on the solutions;
in turn affecting the location of horizons allowing configurations where the mass of the black hole is smaller than its charge. Indeed, this is the case for ModMax as we detail in the next section.

\section{The spherically symmetric charged ModMax black hole}
\label{secc:3}

Let us recall that both Maxwell and Born-Infeld electrodynamics are SO(2) electric-magnetic duality invariant. This means, e.g., that given an electric solution of the theory, by applying a Hodge dualization, we obtain a new magnetic solution to the equations of motion. The constitutive relation might or might not be linear but the equations of motion are. In other words, superposing a solution and its dual leads us to dyonic solutions. For gravity this implies that the most general spherically symmetric black hole is dyonic. This is the case for ModMax dynamics as well. However, to illustrate how certain approaches to this NED are more convenient we study the black hole case by case.

The ModMax field equations are
\be
\nabla_\mu P^{\mu\nu} = 0,
\ee
with $P$ given by \eqref{Pmodmax} and the Einstein equations are
\be
R^\mu{}_\nu- \frac 12 \delta^\mu_\nu R+\Lambda\delta^\mu_\nu = 8\pi T^\mu{}_\nu.
\ee
Just above, the energy momentum tensor $T$ is given by~\cite{Plebanski:1970,Gibbons:2001sx}
\begin{eqnarray}
4\pi T^\mu{}_\nu &=& -F^{\mu\beta} P_{\beta\nu} +  \delta^\mu_\nu L
\nonumber \\[4pt]
&=& F^{\mu\beta} (L_x F_{\beta\nu} + L_y \tilde F_{\beta\nu}) +  \delta^\mu_\nu L,
\nonumber \\[4pt]
&=&\hat L_s P^{\mu\beta} P_{\nu\beta} +  \delta^\mu_\nu (2s \hat L_s + t \hat L_t - \hat L),
\label{tmunu}
\end{eqnarray}
with $\hat L_s := \partial \hat L/\partial s$ and 
$\hat L_t := \partial \hat L/\partial t$.

 Using the Pleba\'nski variables, we may straightforwardly solve the conservation laws; it is known however, that for the electrically charged case, $P_{\mu\nu}$ are the natural variables to use while for the magnetically charged case, the components $F_{\mu\nu}$ are more appropriate~\cite{Ruffini:2013hia}.

Let us now consider a spherically symmetric metric
\be
ds^2 = -f(r) dt^2 + \frac 1{f(r)} dr^2 + r^2 d\Omega,
\label{ds2}
\ee 
where $d\Omega := d\theta^2 + \sin^2\theta d\phi^2$ and $(t,r,\theta,\phi)$ have the usual meanings.
Throughout this section we focus on asymptotically flat configurations and thus choose $\Lambda=0$.
Before writing the field equations in full generality, we give a simple argument leading to the solution for the electrically charged black hole. For this particular case, the electromagnetic invariant $t=0$ and therefore $\hat L_{ModMax} = (\cosh \gamma - \sinh \gamma) s = e^{-\gamma} s = e^{-\gamma} \hat L_{Maxwell}$. Since $\hat L_{Maxwell} = Q_e^2/2r^4$, where $Q_e$ is the charge of the Reissner-Nordström black hole, this results implies that the effective electric charge for the ModMax black hole is $e^{-\gamma/2} Q_e$; in consequence the metric function is simply
\be
f_e(r) = 1 - \frac {2M}r + \frac {Q_e^2 e^{-\gamma}}{r^2},
\label{melec}
\ee
where $M$ is a constant. Obviously, the physical interpretation of $\gamma$ is related to a constant screening factor for the charge of the black hole; although not so obvious, a similar result applies to the magnetically charged black hole. 

To be more clear on the previous statements, in particular for the magnetic case and also to discuss gravitating dyons, we solve the conservation laws $\nabla_\mu P^{\mu\nu} =  \partial_\mu P^{\mu\nu} + \Gamma^\mu_{\mu\sigma} P^{\sigma\nu} = 0$; assuming, as usual, that $P^{\mu\nu} = P^{\mu\nu}(r, \theta)$ and using the metric in Eq.~(\ref{ds2}), we have
\be
\partial_r (r^2 P^{01}) = 0, \qquad \partial_\theta (\sin\theta P^{23}) = 0,
\ee
with solutions
\be
P^{01} = \frac {h_e(\theta)}{r^2}, \qquad P^{23} = \frac {h_m(r)}{r^4 \sin \theta},
\label{pvars}
\ee
where $h_e(\theta), h_m(r)$ are functions to be determined. Notice the presence of $r^4$ in the expression for $P^{23}$; we have chosen it such that $P_{23} = - h_m(r) \sin \theta$ as it is usually done for magnetic solutions. From these solutions, we obtain for the electromagnetic invariants
\be
s = \frac {h_e^2 - h_m^2}{2r^4}, \qquad t = \frac {h_e h_m}{r^4}.
\ee
The Lagrangian $\hat L$ becomes
\be
\hat L_{ModMax} = \frac {h_e^2 e^{-\gamma} - h_m^2 e^\gamma}{2r^4}.
\ee
In the following, for convenience purposes, we set $f = 1 - 2 m(r)/r$ for the metric coefficient.

\subsection{Electrically charged ModMax black hole}

For this solution, $h_e = -Q_e = const.$, and $h_m=0$; the only non-vanishing Pleba\'nski variables are then
\be
P_{\mu\nu} = \frac {Q_e}{r^2} \,\delta^0_{[\mu} \delta^1_{\nu]}.
\ee
It follows that $s = Q_e^2 e^{-\gamma}/2r^4, t = 0$. Using the third line in Eq.~(\ref{tmunu}) for $T^\mu{}_\nu$, we find that the solution to the field equations 
\begin{eqnarray}
-\frac {m_{,r}}{r^2} = -\frac {Q_e^2 e^{-\gamma}}{2r^4}, \qquad 
-\frac {m_{,rr}}{2r} = \frac {Q_e^2 e^{-\gamma}}{2r^4},
\end{eqnarray}
is 
$$
m(r) = M - \frac {Q_e^2 e^{-\gamma}}{2r},
$$
where $M$ is an integration constant; thus, we recover Eq.~(\ref{melec})

\subsection{Magnetically charged ModMax black hole}
\label{secc:3.2}
As mentioned before, the suitable variable for this case is the electromagnetic tensor $F_{\mu\nu}$. We take thus
\be
F_{\mu\nu} = -Q_m \cos \theta \, \delta^2_{[\mu} \delta^3_{\nu]}.
\ee
It follows that $x = Q_m^2 e^{-\gamma}/2r^4, y = 0$. Furthermore, we have
\be
P_{01} = 0, \quad P_{23} = -L_x F_{23} = - Q_m e^{-\gamma} \sin \theta,
\ee
ensuring that the conservation laws $\nabla_\mu P^{\mu\nu} = 0$ are satisfied. Using the second line in Eq.~(\ref{tmunu}) for $T^\mu{}_\nu$, the field equations become
\begin{eqnarray}
-\frac {m_{,r}}{r^2} = -\frac {Q_m^2 e^{-\gamma}}{2r^4}, \qquad 
-\frac {m_{,rr}}{2r} = \frac {Q_m^2 e^{-\gamma}}{2r^4},
\end{eqnarray}
with solution
$$
m(r) = M - \frac {Q_m^2 e^{-\gamma}}{2r},
$$
where $M$ is again an integration constant; thus, we obtain the expression
\be
f_m(r) = 1 - \frac {2M}r + \frac {Q_m^2 e^{-\gamma}}{r^2},
\label{mmag}
\ee
for the metric function.

Related to the magnetic solution just found, and recalling Eqs.~(\ref{pvars}), we may also look at the situation when $h_e = 0, h_m = \bar Q_m = const.$; in this case $P_{23} = - \bar Q_m \sin \theta$. The field equations become then
\begin{eqnarray}
-\frac {m_{,r}}{r^2} = -\frac {\bar Q_m^2 e^\gamma}{2r^4}, \qquad
-\frac {m_{,rr}}{2r} = \frac {\bar Q_m^2 e^\gamma}{2r^4}.
\end{eqnarray}
It is clear by now that 
\be
m(r) = M - \frac {\bar Q_m^2 e^\gamma}{2r},
\ee
and that the metric function is
\be
f_m(r) = 1 - \frac {2M}r + \frac {\bar Q_m^2 e^\gamma}{r^2}.
\label{mmag2}
\ee
At first sight, we obtained a new magnetic solution with a screening factor $e^\gamma$ for the magnetic charge. Nevertheless, if we recall that $P_{23} = -Q_m e^{-\gamma} \cos \theta$ in the magnetic case discussed previously, we make the identification $\bar Q_m = Q_m e^{-\gamma}$. Substitution of this relation in Eq.~(\ref{mmag2}) gives us back the metric in Eq.~(\ref{mmag}); basically in this point of view, the charge $Q_m$ is more fundamental than $\bar Q_m$.

\subsection{Dyonic ModMax black hole}

To analyse dyon solutions, we take first $A_\mu = (\Phi(r), 0 , 0, Q_m\cos \theta)$. It follows that
\be
F_{\mu\nu} = - \Phi_{,r} \delta^0_{[\mu} \delta^1_{\nu]} - Q_m \sin \theta \delta^2_{[\mu} \delta^3_{\nu]}.
\ee
In consequence
\begin{eqnarray}
&&x = \frac 12 \left( -\Phi_{,r}^2 + \frac {Q_m^2}{r^4} \right), \qquad y = - \frac {\Phi_{,r} Q_m}{r^2},
\nonumber \\[4pt]
&&(x^2 + y^2)^{1/2} = \frac 12 \left( \Phi_{,r}^2 + \frac {Q_m^2}{r^4} \right),
\end{eqnarray}
and after some algebra, the only non-vanishing Pleba\'nski variables are
\begin{align}
P_{01} &= -(\cosh \gamma + \sinh \gamma) \Phi_{,r} = -e^\gamma \Phi_{,r}, 
\nonumber \\[4pt]
P_{23} &= -(\cosh \gamma - \sinh \gamma) Q_m \sin \theta = - e^{-\gamma} Q_m \sin \theta.
\end{align}
The expression for $P_{23}$ automatically satisfies one of the conservation equations. For the electric potential, we write 
\be
\Phi(r) = \frac {Q_e e^{-\gamma}}r, 
\ee
implying that the conservation law associated to $P_{01}$ also holds; furthermore, we have $P_{01} = Q_e/r^2$. With these choices, the gravitational field equations read
\begin{eqnarray}
-\frac {m_{,r}}{r^2} &=& -\frac {(Q_e^2 + Q_m^2) e^{-\gamma}}{2r^4},
\nonumber \\[4pt]
-\frac {m_{,rr}}{2r} &=& \frac {(Q_e^2 + Q_m^2) e^{-\gamma}}{2r^4}.
\end{eqnarray}
It follows that
\be
m(r) = M - \frac {(Q_e^2 + Q_m^2)e^{-\gamma}}{2r},
\ee
and in consequence, the metric coefficient is 
\be
f_{dyon}(r) = 1 - \frac {2M}r + \frac {(Q_e^2 + Q_m^2)e^{-\gamma}}{r^2}.
\ee

An important feature of the electromagnetic solutions discussed previously concerns the presence of horizons; we have for example 
\be
r_\pm = M \pm \sqrt{M^2 - Q_e^2 e^{-\gamma}},
\ee 
for the electrically charged ModMax black hole. As it is well-known, for the RN black hole the mass must be always greater than the charge; for the ModMax black hole this is not necessarily the case. Indeed, the extremal configuration for the electrically ModMax black hole, where $r_+ = r_-$, implies
\be
M_{ext} = |Q_e| e^{-\gamma/2}.
\ee
It follows that if $\gamma > 0$, then $M_{ext} < |Q_e|$; similar results are known for other non-linear electrodynamics~\cite{Ruffini:2013hia,Maceda:2018zim}.

To close this section, we mention that, as usual, in the near-horizon limit a Bertotti-Robinson-like spacetime is obtained from the spherically symmetric ModMax black hole.

\section{Exact gravitational waves}
\label{secc:4}

We now focus on explicit exact gravitational wave solutions to Einstein theory coupled to ModMax electrodynamics. We use coordinates $(u,v,x,y)$
throughout. The directions associated with $u$ and $v$
are null, they are interpreted as retarded and advanced time, respectively.

Thus far, we have illustrated how some standard electrovacuum configurations easily make their way into 
the ModMax scenario. These configurations, e.g., those with $y=0$, greatly simplify the equations of motion.
However, the nonlinear nature of the electrodynamics
has unexpected effects on the configuration.

To further explore this point, let us consider, the exceptional Pleba\'nski-Hacyan universe~\cite{Plebanski:1979}. This electrovacuum solution requires a nonzero cosmological constant, it has a geometry given by
\begin{equation}
 ds^2 = 2\Lambda v^2du^2-2dvdu + dx^2+dy^2,
\end{equation}
and harbors a field
\begin{equation}
F_{\mu\nu} = Q_m\, \delta^2_{[\mu} \delta^3_{\nu]}.
\label{FPH}
\end{equation}
Proceeding as in Sec.~\ref{secc:3.2} we first note that 
$x=2Q_m^2$ and $y=0$. Thus, the equations of motion
fix the cosmological constant by
\begin{equation}
 \Lambda=-e^{-\gamma}Q_m^2.
\end{equation}
Once again, we see charge screening in the geometry due to the interaction with a nonlinear electrodynamics.
Notice that when $\gamma$ is arbitrarily large then 
the cosmological constant arbitrarily approaches zero and the geometry becomes ever so closer to Minkowski spacetime. However, the flux density $F$ is unaffected by this. This undesireable feature is not problematic as phenomenology expects $\gamma$ to be small.

Another feature of this spacetime is that it is able to support exact gravitational waves. Extending the metric by a Kerr-Schild-like term~\cite{Kerr:1965} yields
\begin{equation}
 ds^2 = (2\Lambda v^2-H)du^2-2dvdu + dx^2+dy^2,
 \label{typeII}
\end{equation}
where $H=H(u,x,y)$ is an arbitrary function of $u$
but is a harmonic function in the $(x,y)$-plane.
The coordinate $u$ plays the r\^ole of retarded time and $H$ of a gravitational wave profile.
Geometrically, this spacetime is of Kundt type and algebraically Petrov type II. Thus, it represents type II gravitational waves over the Pleba\'nski-Hacyan universe~\cite{Podolsky:2002sy}. Hence, we have provided a class of exact gravitational waves in
General Relativity coupled to ModMax electrodynamics
given by equations \eqref{FPH} through \eqref{typeII}.

At the same time, it seems natural to explore pp-waves in gravitating ModMax configurations, as a distinctive attribute of this NED is that it possesses plane-wave solutions in the interaction regime (as long as $\gamma>0$). Gravitating electromagnetic plane-waves belong to the conformally flat family of solutions introduced by Baldwin-Jeffery and 
Brdi\v{c}ka~\cite{Baldwin:1926,Brdicka:1951}.
In the light of reference~\cite{Podolsky:2002sy}, we propose an Ansatz with line element
\begin{equation}
 ds^2=[(x^2+y^2)A-H]du^2-2dvdu + dx^2+dy^2, \label{typeN}
\end{equation}
and null field
\begin{equation}
 F=f(u)\, \left(\cos\zeta\delta^1_{[\mu} \delta^3_{\nu]}+\sin\zeta\delta^1_{[\mu} \delta^4_{\nu]}\right), \label{FBJB}
\end{equation}
where $H=H(u,x,y)$ is as above and $A=A(u)$ is to be fixed by the matter source. In the previous equation $f(u)$ is the plane-wave profile and $\zeta$ an angle which determines the propagation direction.
This configuration is a solution of GR-ModMax as long as
\begin{equation}
 A(u)=\cosh\gamma f(u)^2. \label{ppwaves}
\end{equation}
In comparison with Maxwellian matter, there is a stronger back-reaction of plane-waves on the background for ModMax.
This class of Kundt waves is rather different to the one given above. The family of configurations given by equations \eqref{typeN}-\eqref{ppwaves} represent conformally flat pp-waves sourced by pure electromagnetic radiation. While the type II waves of
\eqref{FPH}-\eqref{typeII} are sourced by matter absent of pure radiation.

\section{Light-cone structure and birefringence}
\label{secc:5}

To discuss birefringence, we first remember that Maxwell's equations in curved spacetime can be put into Faraday form by introducing appropriate dielectric permittivity and magnetic permeability tensors~\cite{Balakin:1997dv}; this approach can be extended by considering effective metrics and generalised Fresnel equation~\cite{Plebanski:1970,Hehl:2004yk,Lammerzahl:2004ww}. For non-linear electrodynamics, we follow the treatment in~\cite{Obukhov:2002xa} where the covariant Fresnel equation
\be
{\cal G}^{\mu\nu\sigma\rho} q_\mu q_\nu q_\sigma q_\rho = 0
\ee
allow to determine if different light cones exist for wave normal covectors. The analysis in~\cite{Obukhov:2002xa} shows that two optical metrics $g^{\mu\nu}_{1,2}$ arise such that the Fresnel equation becomes
\be
(g_1^{\mu\nu} q_\mu q_\nu) (g_2^{\sigma\rho} q_\sigma q_\rho) = 0.
\ee
In the case of ModMax non-linear electrodynamics, we verify straightforwardly that the necessary and sufficient conditions~\cite{Obukhov:2002xa} for the absence of birefringence do not hold; indeed, we obtain for the optical metrics
\begin{eqnarray}
g_1^{\mu\nu} &=& \frac{x^2 + y^2 \cosh^2 \gamma}{x^2 + y^2} g^{\mu\nu} 
\nonumber \\[4pt]
&&+ 16 \sinh \gamma \frac {x \sinh \gamma - (x^2 + y^2)^{1/2} \cosh \gamma}{x^2 + y^2} t^{\mu\nu},
\nonumber \\[4pt]
g_2^{\mu\nu} &=& \frac{x^2 + y^2 \cosh^2 \gamma}{x^2 + y^2} g^{\mu\nu}, 
\end{eqnarray}
where in general
\begin{align}
t^{\mu\nu} &:= F^{\mu\sigma} F^{\nu}{}_\sigma 
\nonumber \\[4pt]
&= diag(-g^{00} \Phi_{,r}^2, -g^{11} \Phi_{,r}^2, g^{22} \frac {Q_m^2}{r^4}, g^{33} \frac {Q_m^2}{r^4}).
\end{align}
The above expressions for the optical metrics show that we have birefringence with some photons following trajectories determined by the standard geometric metric. Only when $\gamma = 0$, we recover propagation of photons along a unique light-cone; we arrive to the same conclusion employing other formalisms~\cite{Bialynicki:1983,deMelo:2014isa}. 

\section{Conclusions}

The ModMax black hole exhibits a feature shared by other non-linear electrodynamics, namely, a screening factor depending on the non-linearity parameter $\gamma$ that shields the actual charges. The screening factor has a direct influence on the horizons; we see explicitly that the location of the horizons changes when compared with the standard Reissner-Nordström black hole. As a consequence of the screening, a ModMax black hole admits horizons where its mass is smaller than its charge; this feature is more noticeable in the extremal case, where there is only one horizon and the ratio between mass and charge is simply $e^{-\gamma/2}$. 

Gravitational waves are shown to exist for ModMax matter of both nonradiating and purely radiating nature; with and without cosmological constant, respectively. These include pp-waves sourced by plane-waves in the interacting regime $\gamma>0$. In this case ModMax plane-waves display a stronger back-reaction on the background than their Maxwell equivalent.

We also showed that birefringence exists, even if the ModMax metric has a Reissner-Nordström like form. Since the optical metrics were found, with one of the them differing from the geometric metric, we may analyse the trajectories of null rays in the effective metric and compare them with those associated to the standard spacetime metric. It would be also interesting to set constraints on the non-linearity parameter $\gamma$ along the lines in~\cite{Gleiser:2001rm} but that analysis is beyond the scope of the present work.

\begin{figure}[htbp]
\begin{center}
\includegraphics[width=12cm]{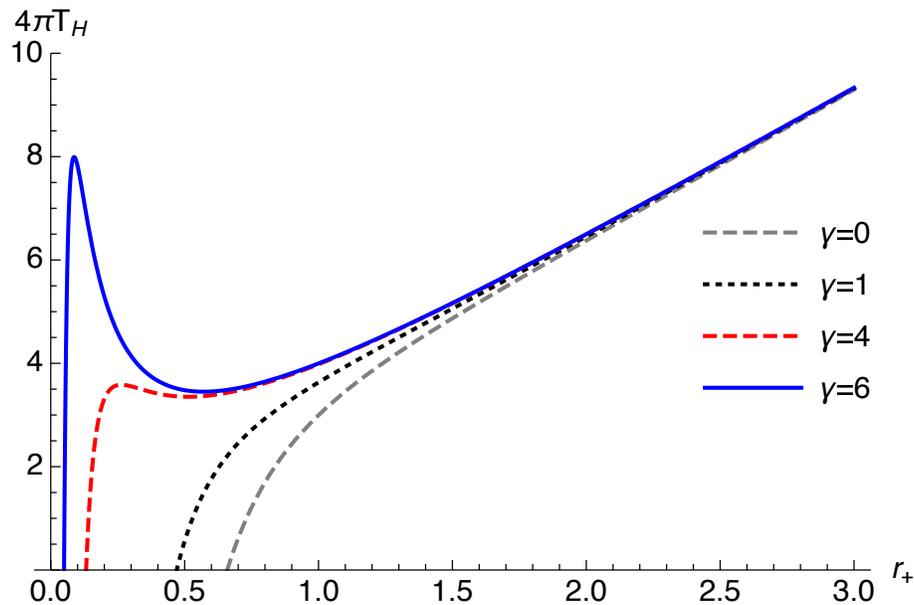}
\caption{The Hawking temperature $T_H$ of the electrically charged anti-de Sitter-ModMax black hole as a function of the (outer) horizon radius $r_+$ for different positive values of the non-linearity parameter $\gamma$ with $|Q_e| = 1$ and $l=1$; the dotted curve corresponds to the Reissner-Nordström black hole ($\gamma = 0$). There is a value $r_+ = \bar r_+$ where $T_H$ vanishes; (local) extrema of $T_H$ appear if $6 |Q_e| e^{-\gamma/2} < l$ holds.}
\label{fig1}
\end{center}
\end{figure}

Future work to pursue include the thermodynamical properties of the ModMax black hole. As a starter, we may add to the field equations a cosmological term $-3 l^{-2} g_{\mu\nu}$ to calculate the Hawking temperature of the anti-de Sitter-ModMax black hole; for the electrical case, a straightforward calculation gives
\be
T_H = \frac 1{4\pi r_+} \left(1 - \frac {Q_e^2 e^{-\gamma}}{r_+^2} + 3\frac {r_+^2}{l^2} \right).
\ee
In Fig.~\ref{fig1} we show the dependence of $T_H$ as a function of the (outer) horizon radius $r_+$; for definiteness, we consider $\gamma \geq 0$ in the following. The Hawking temperature $T_H$ vanishes at
\be
\bar r_+ = \frac {\sqrt{2} |Q_e| e^{-\gamma/2}}{\sqrt{1 + \sqrt{1 + 12 Q_e^2 e^{-\gamma} l^{-2}}}}, 
\ee
indicating the presence of an extremal black hole; we see that in the anti-de Sitter-ModMax black hole, $\bar r_+$ goes to zero as $\gamma$ increases. We also notice that two local extrema for $T_H$ appear; they are located at 
\be
r_{+\,1,2} = \frac {\sqrt{6} |Q_e| e^{-\gamma/2}}{\sqrt{1 \pm \sqrt{1 - 36 Q_e^2 e^{-\gamma} l^{-2}}}}, 
\ee
provided $6 |Q_e| e^{-\gamma/2} < l$, with $r_{+\,1}$ associated to a maximum and $r_{+\,2}$ to a minimum. As $\gamma$ increases, $r_{+\,1} \sim \sqrt{3} |Q_e| e^{-\gamma/2} \to 0$ while $r_{+\,2} \to l/\sqrt{3}$; furthermore, in the limit $l \to \infty$, we have the relation $\bar r_+ = \sqrt{3} r_{+\,1}$ for fixed $\gamma$. For large values of $r_+$, the Hawking temperature of the anti-de Sitter-ModMax black hole becomes indistinguishable of the temperature of the anti-de Sitter-Schwarzschild black hole, just as it happens for the anti-de Sitter-Reissner-Nordström black hole. A more extensive analysis of the thermodynamical properties of this spacetime will be discussed elsewhere. Thus, we expect a phase structure similar to charged anti-de Sitter black holes~\cite{Chamblin:1999tk}.

\section*{Acknowledgements}

DFA thanks the Mexican Secretariat of Public Education(Secretar\'ia de Educaci\'on P\'ublica) for support through grant PRODEP No. 12313509.


\end{document}